\documentclass[preprint,showpacs,preprintnumbers,amsmath,amssymb,a4paper,endfloats*]{revtex4}

\usepackage{graphicx}
\usepackage{psfrag}
\usepackage{subfigure}
\usepackage[a4paper,dvips]{geometry}
\usepackage{here}

\begin{document}

\title{Dual view FIDA measurements on MAST}

\author{C.A. Michael$^1$, N. Conway, B. Crowley, O. Jones$^2$, W.W. Heidbrink$^3$, S. Pinches$^4$, E. Braeken$^5$, R. Akers, C. Challis, M. Turnyanskiy, A. Patel, D. Muir, R. Gaffka and S. Bailey}

\address{EURATOM/CCFE Fusion Association. Culham Science Centre, Abingdon, Oxon, OX14 3DB, UK}
\address{$^1$ Australian National University, Canberra, A.C.T. 0200}
\address{$^2$ Durham University, Durham, United Kingdom}
\address{$^3$ University of California Irvine, USA}
\address{$^4$ ITER Organization, Route de Vinon-sur-Verdon, 13115, St. Paul-lez-Durance, France}
\address{$^5$ Eindhoven University, Netherlands}
\date{\today}
\begin{abstract}
A Fast Ion Deuterium Alpha (FIDA) spectrometer was installed on MAST
to measure radially resolved information about the fast ion density
and its distribution in energy and pitch angle. Toroidally and
vertically-directed collection lenses are employed, to detect both
passing and trapped particle dynamics, and reference views are
installed to subtract the background. This background is found to
contain a substantial amount of passive FIDA emission driven by edge
neutrals, and to depend delicately on viewing geometry.  Results are
compared with theoretical expectations based on the codes NUBEAM (for
fast ion distributions) and FIDASIM.  Calibrating via the measured
beam emission peaks, the toroidal FIDA signal profile agrees with
classical simulations in MHD quiescent discharges where the neutron
rate is also classical.  Long-lived modes (LLM) and chirping modes
decrease the core FIDA signal significantly, and the profile can be
matched closely to simulations using anomalous diffusive transport; a
spatially uniform diffusion coefficient is sufficient for chirping
modes, while a core localized diffusion is better for a LLM.  Analysis
of a discharge with chirping mode activity shows a dramatic drop in
the core FIDA signal and rapid increase in the edge passive signal at
the onset of the burst indicating a very rapid redistribution towards
the edge.  Vertical viewing measurements show a discrepancy with
simulations at higher Doppler shifts when the neutron rate is
classical, which, combined with the fact that the toroidal signals
agree, means that the difference must be occurring for pitch angles
near the trapped-passing boundary.  Further evidence of an anomalous
transport mechanism for these particles is provided by the fact that
an increase of beam power does not increase the higher energy vertical
FIDA signals, while the toroidal signals do increase.
\end{abstract} \maketitle
\section{Introduction}

The physics of fast ions is of particular importance from the point of
view of fundamental understanding for predicting the performance of
burning plasmas, as well as making correct interpretation of data on
present day plasma experiments where energetic ions play a role such
as in neutral beam heating and ion cyclotron resonance heating.  The
classical collisional slowing down process
\cite{corecordey1974,core1993} is the primary effect required to
describe the fast ion distribution function.  However, instabilities
can also alter the distribution function.  Large fast-ion transport by
fishbones \cite{MCGUIRE83A}, toroidal Alfv\'en eigenmodes (TAEs)
\cite{DUONG93B}, and neoclassical tearing modes (NTMs) \cite{CAROLIPIO02A}
have been measured; other modes that could modify the distribution
function include
 magnetic micro-turbulence \cite{hauff},
 the ideal infernal mode \cite{iansat}, and
compressional Alfv\'en eigenmodes (CAEs) \cite{FREDRICKSON01A}.
 Because the magnetic field is low in spherical
tokamaks, injected fast ions are super-Alfv\'enic so, during the
slowing down process,
 many resonances that can cause transport can occur. 
For example, in MAST,  the
toroidal transit frequency of full-energy beam ions is $\sim 300$kHz, while
the TAE frequency is $\sim 100$kHz and the fishbone
frequency is even lower ($\sim 30-50$kHz).
Four mechanisms of resonant fast-ion transport have been identified
\cite{heidbrinkreview}. With global modes such as the fishbone, 
phase-locked convective transport occurs when the particles stay in phase
with the wave as they steadily march out of the machine
\cite{WHITE83A}. Losses can also occur when the wave-particle interaction
causes a change in orbit topology and places the fast ion onto an
unconfined banana orbit \cite{SIGMAR92A}. In the presence of many
small-amplitude resonances, diffusive transport occurs
\cite{WHITE10A}. A fourth type of transport is analogous to the
collapse of a sandpile avalanche,
where steepening of the profile by one mode causes progressive
destabilization of modes at other spatial locations \cite{BERK95C}.
In addition to these resonant processes,
the orbits of sufficiently energetic fast ions can become stochastic
in the presence of a large helical field perturbation introduced by
an NTM or infernal mode \cite{KONOVALOV88A,MYNICK93A}.  Recent experiments
have also demonstrated that, in the presence
of loss boundaries, non-resonant interactions of fast ions with
Alfv\'en eigenmodes can cause substantial losses \cite{CHENXI13A}.
For all of these processes, the effect on the distribution function
ultimately depends on the amplitudes of the instabilities. The
  nonlinear processes that determine these amplitudes are
difficult to model accurately, 
so experiments are necessary to evaluate
which processes are most important.

The Mega-Amp Spherical Tokamak features two neutral beam injectors
(NBIs) of up to 75keV energy and
2.5MW each \cite{gee}. These injectors are separated toroidally by $60^{\circ}$ and are labelled SS and SW.  A variety of fast
ion driven MHD such as fishbones \cite{iansat} and TAEs \cite{miktae} has been reported based on magnetic measurements. Fast ion
diagnostics have included a compact neutral particle analyzer (CNPA) and
a main NPA \cite{npamast}.  The main absolutely-calibrated diagnostic
is the fission chamber \cite{mastfission}. It has been shown
\cite{mikaxis} that the measured neutron rate is lower than the rate
expected based on collisional classical slowing down, particularly at
higher power. The measured neutron rate also drops after fishbone
activity. A counter-viewing bolometer is sensitive to lost fast
ions and shows spikes during fishbone events \cite{iansat}. Transport
studies \cite{anthonytransport,valovic} have also been hampered by
accuracy of the ion power source due to fast ion slowing down. Using
the NUBEAM code \cite{transp}, anomalous fast ion diffusion can be
introduced into the modelling. There are many possible ways to include this,
as the spatial and energy dependencies can be varied and convection may also be included. In \cite{anthonytransport}, using an
anomalous diffusion of $2\mathrm{m}^2/\mathrm{s}$ reduced the
predicted neutron rate to the measured value, while other
parameterisations which maintain the stored energy and Shafranov shift
(which depend on fast ion pressure) result in a negative thermal
conductivity, indicating the need for improved fast ion diagnostics to
support transport analysis.  The neutron rate, EFIT
\cite{lyntonefitstart} derived stored energy and Shafranov shift are
somewhat blunt instruments in the sense that they are not local
quantities. The MSE-constrained EFIT-derived total pressure, which includes a fast ion component, is
unreliable as it often depends on the data weights. In future, the use of
Bayesian methods may provide a reliability check on the equilibrium
derived pressure \cite{holebayesian}, but for now better
quantification of the fast ion losses through profile measurements is
essential and has led to the recent development of the FIDA diagnostic
as well as a neutron camera system \cite{neutcamera}. Complementary
analysis of all these fast ion diagnostics will improve
understanding of the fast ion physics as different diagnostics have
different degrees of sensitivity in energy/pitch and real space
\cite{heidbrinkweight}.

The fast ion $\mathrm{D}_{\alpha}$ (FIDA) spectroscopy technique
\cite{heidbrinkreview} is an appealing way to diagnose local and
energy-resolved information about the fast ion distribution function,
which is essential to measure spatial redistribution, where
spatially-integrated measurements such as a fission chamber neutron
counter are less sensitive. The first application of a charge
exchange diagnostic to infer details of the fast ion distribution used
He spectroscopy on JET \cite{vonhellerman}, while the first true
FIDA experiments were carried out on DIII-D \cite{heidbrinkorig}.
Since then the technique has been applied to various machines
\cite{asdexfida,lhdfida,textorfida,fidavertnstx,fidatannstx}.
Modelling of FIDA signals has advanced over the years with the
development of the FIDASIM code \cite{fidasim}. In many machines,
FIDA measurements have been made by re-configuring the CXRS optics. Pilot experiments to test the FIDA concept on MAST were carried out
during late 2009 using both the main CXRS system in MAST
\cite{mastcxrs}, as well as an auxiliary high resolution spectrometer
and fibres dedicated for MSE \cite{mastmse}. The results were
encouraging but not of high quality because of the presence of a very bright $\mathrm{D}_{\alpha}$ emission peak and the lack
of flexibility of the CXRS spectrometer. The FIDA system
described here was therefore installed in 2010.

The purpose of this paper is to describe in detail the instrument on MAST, as well as to show preliminary results demonstrating the potential of the diagnostic to provide new information about the fast ion distribution.

This paper is organized as follows.  In section II, the FIDA
installation on MAST, including sight lines and spectrometer, is
described. In Section III, calibration and validation are considered
(essential to properly understand whether the fast ion signal obeys a
classical distribution) and the reliability of the different
background subtraction methods is compared. In section IV,
simulations are described which are used for comparison with
measurements in section V. Results are analysed in an MHD quiescent,
`classical' discharge, for both the toroidal (passing) and vertical
(near the trapped/passing boundary) signals. The toroidal signals are
analyzed in a discharge with a steady LLM, and a discharge with
chirping mode activity, including a comparison of the time history to 
`limit cycle' behaviour.

\section{Diagnostic setup}

\subsection{Sight lines}

\begin{figure}[htb]
\includegraphics[width=8cm]{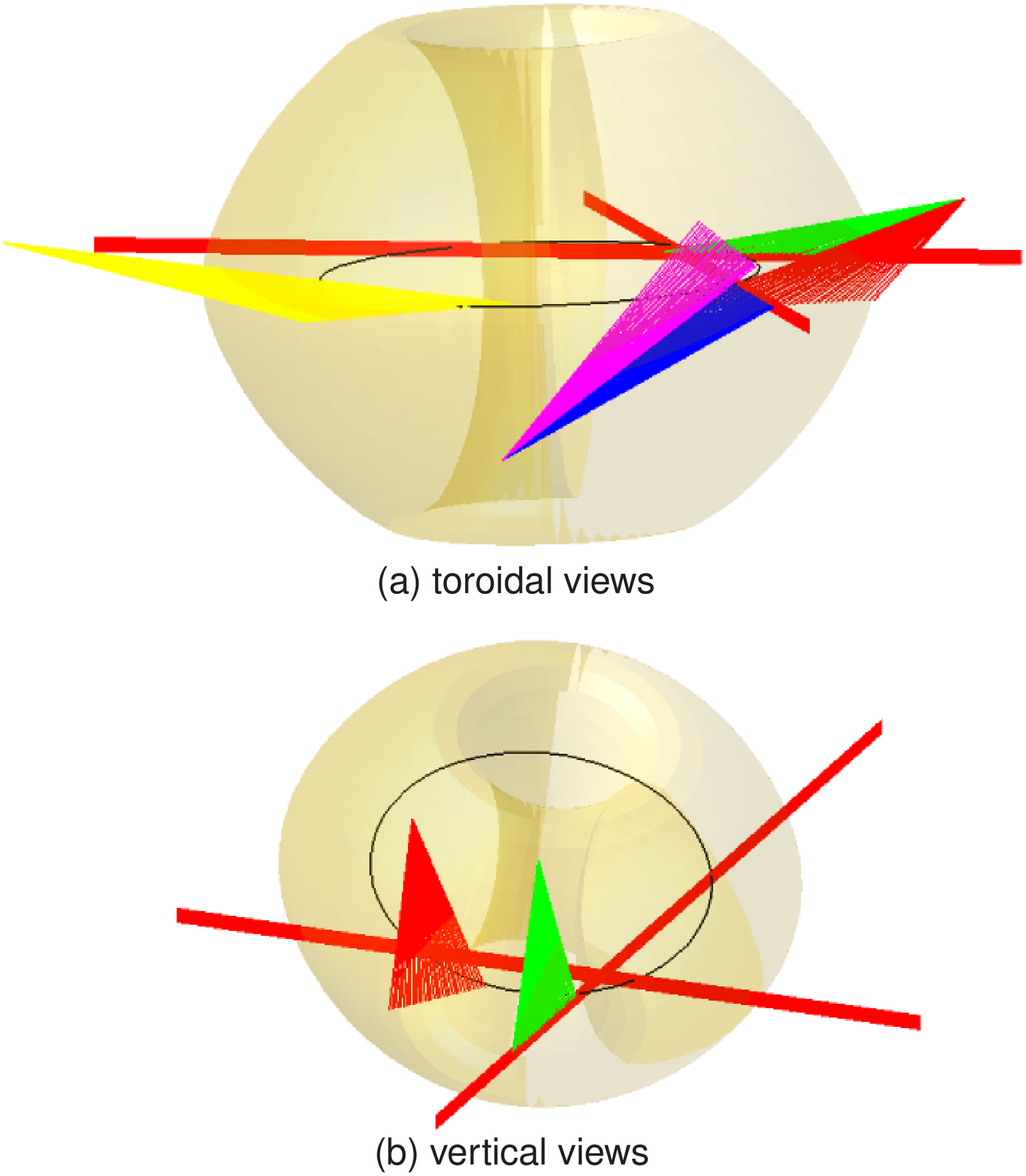}
\caption{3D view of the sight lines of (a) toroidally-viewing chord bundles (blue: active on the SS beam, green: active on the SW beam, purple: vertically-displaced reference on the SW beam, red: vertically-displaced reference on the SS beam, yellow: toroidally-displaced reference) and (b) vertically-viewing chord bundles (green: active on the SS beam and red: toroidally-displaced reference view).}
\label{fig:chords}
\end{figure}

Owing to limited port access it was decided to utilize the CXRS viewing
optics \cite{mastcxrs}, as these already include toroidal (tangential)
and vertical (perpendicular) viewing fibres for toroidal and poloidal
rotation measurements.  Also, the toroidal views have an approximately
$8^{\circ}$ vertical tilt which makes the sight lines pass mostly
through a single beam rather than both beams, thereby allowing better
spatial localization. For the toroidal views, a reference view is
installed $90^{\circ}$ around from the closest beam in order to
subtract bremsstrahlung and passive FIDA components. The toroidal
views are shown in Figure (\ref{fig:chords}a). FIDA was combined with
the CXRS system by adding a second row of fibres to the collection
lens, such that the imaged spots on the beam (approximately 1.5cm diameter) from backlighting the CXRS and FIDA fibres
were coincident with one another. In addition, to provide contingency
against toroidal asymmetries in the background emission (which can
arise from gas puffing, for example \cite{marcothesis,asdexfida}),
additional chords were installed which view approximately the same
radius as the beam viewing chords but are vertically displaced. This
was achieved by having a third row of fibres inserted into the active
viewing lens, to collect light from 25cm above or below the
beam (depending on rotation of the lens assembly). There are 32 chords available on the active view from
$R=0.77-1.4\mathrm{m}$ and 32 chords on each of the toroidally and
vertically-displaced reference views. While the CXRS system has views
on both SS and SW beams, FIDA views were only installed on a single
beam. The sight lines were interchanged between different
beams during the MAST M8 experimental campaign. The vertically-displaced reference views are placed on the side of the beam which
passes though a greater path length of plasma and a similar flux
coordinate at the edge, to ensure that the background
component is similar on active and reference views.

Vertical views were also installed. Direct vertical viewing access is
difficult in MAST because of the divertor plates. Previously, the
CXRS installation \cite{mastcxrs, mastpoloidal} was such that there
were 4 vertical views - internally mounted lens coupled fibres
directed to an optical vacuum feed through - consisting of a pair of
terminated 8x8 fibre arrays on either side of a window, together with
imaging optics.  There were in total an active and reference view on
both the SS and SW beams. However, the CXRS spectrometer could only
utilize one beam at a time. This gave the opportunity of utilizing
the unused views for FIDA. New fibres were installed to take light from
the internal SS beam-viewing active and reference fibre bundles to a
patch panel, and from there to the spectrometer. The active viewing lenses were relocated such that each chord passed through the centreline of the beam. The reference view was relocated toroidally by around $30^{\circ}$ such that it
avoided both the SS and SW beams. The achieved geometry is shown in
Figure (\ref{fig:chords}b).  There are 32 active and 32 reference
vertical fibres spanning a region from 1.0-1.3m, limited on the
inner and outer sides by vertical and resonant magnetic perturbation
field coils respectively.

\subsection{Response functions and spectral shape}

\begin{figure}[htb]
\includegraphics[width=8cm]{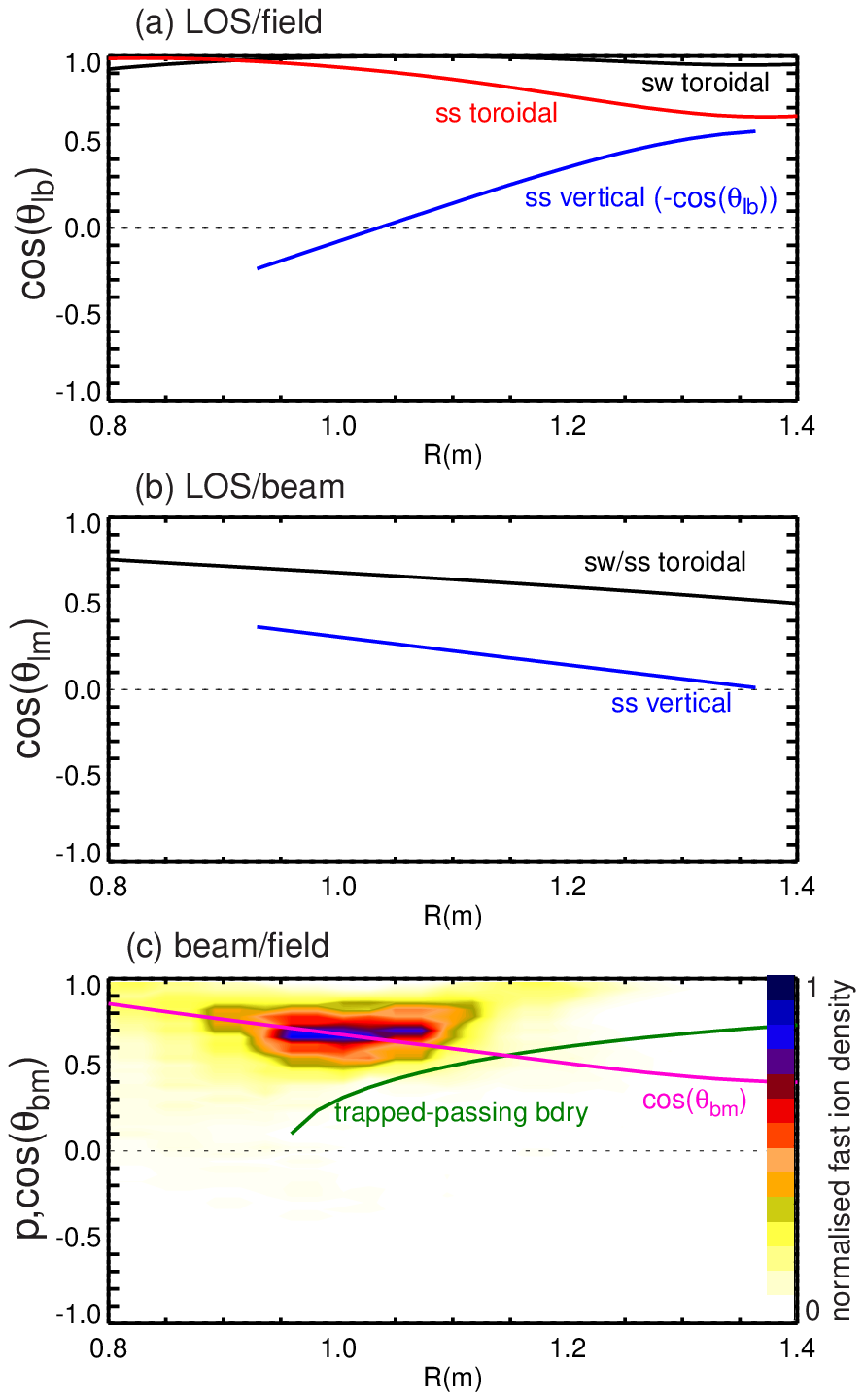}
\caption{(a) Cosine of angle between the line of sight and the
  magnetic field at the position of the beam for different views (the
  value on the SS vertical is inverted to compare blue shifted
  wavelengths with red shifted ones with the same value of $E_{\rm
  min}$). (b) Cosine of angle between the beam and the line of sight,
  which determines the beam emission wavelengths. (c) Simulated TRANSP
  fast ion distribution (normalised) for MAST shot $\#26887$ for $E>30keV$ as a
  function of $(R,p)$ and birth pitch parameter $\cos
  \theta_{bm}$. (For this shot, $B_t = 0.5$T, $I_p = 800$kA.) The
  trapped-passing boundary \cite{orbitbook} is shown for reference.}
\label{fig:dotp}
\end{figure}

The viewing direction defines the range of pitch angles in the
distribution function to which the diagnostic is sensitive, according
to the algebra given in the appendix of \cite{heidbrinkweight}. It can
be shown that the maximum possible Doppler shift from a reneutralized
fast ion gyrating with energy $E$ and pitch parameter
$p=v_{\parallel}/v$ is given by \cite{mirko}
\begin{equation}
\sqrt{\frac{E_{\min}}{E}} = p \cos \theta_{lb}  + \sqrt{(1-\cos^2 \theta_{lb})(1-p^2)},
\end{equation}
where $E_{\min} = m c^2 (\Delta \lambda)^2/2 \lambda^2$ with $\lambda$
and $\Delta \lambda$ being the unshifted and Doppler shifted
wavelengths respectively, and $\theta_{lb}$ is the angle between the
line of sight $l$ and the magnetic field $b$. It can furthermore be
shown that $p=\cos \theta_{lb}$ is the pitch parameter at which E is
minimised, however most of the signal has contributions from higher
energies at different pitch angles. The spatial profiles of
$\cos\theta_{lb}$ are given in Figure (\ref{fig:dotp}a), for all the
possible active views (SS/SW toroidal and SS vertical). The SW
toroidal view, as it is inclined upward, is almost entirely tangential
to the field; the SS toroidal view is inclined downward, resulting in
lower values of $\cos \theta_{lb}$. The SS vertical view is completely
perpendicular to the field near the magnetic axis ($\sim 0.9$m) having
$-\cos\theta_{lb}=0$, rising up to $-\cos\theta_{lb}=0.6$ near the
edge (the negative sign used because the relevant spectral region is
blue shifted).  Therefore, towards the core, there is a roughly equal
distribution of blue and red-shifted light, while towards the edge
there is more blue shift.

\begin{figure}[htb]
\includegraphics[width=8cm]{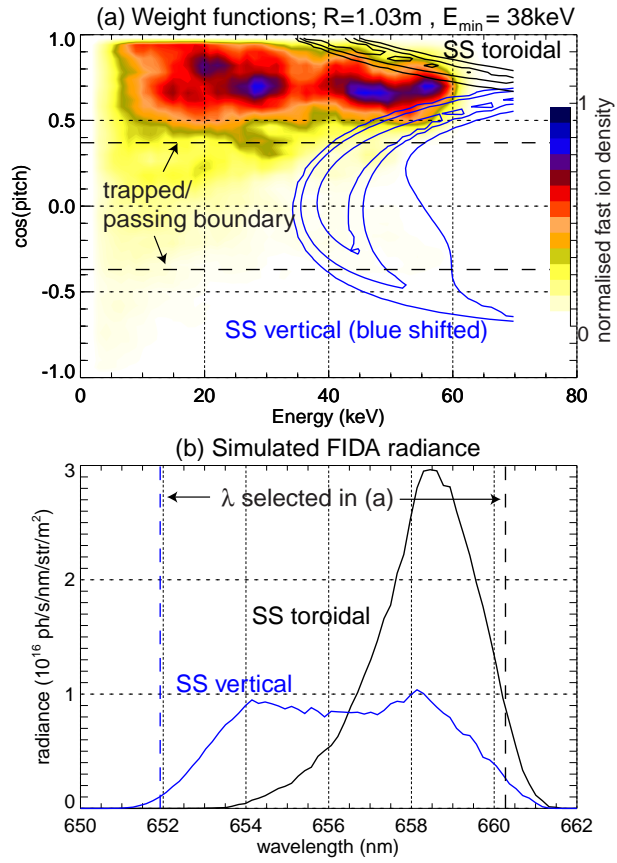}
\caption{(a) The TRANSP-simulated fast ion distribution (for shot $\#26887$, $t=0.25$s) at $R=1.03$m as a function of pitch and energy together with the approximate energy response functions at $E_{\min}=38keV$ for the SS toroidal view ($\lambda=660$nm) and vertical view ($\lambda=652$nm). Horizontal dashed lines indicate the trapped/passing boundary. (b) FIDA radiance for the horizontal and vertical chords looking at $R=1.02$m derived from the FIDASIM code for the same shot and time, indicating the wavelengths corresponding to the weight functions plotted in (a).}
\label{fig:simfida}
\end{figure}

The energy and pitch parameter resolved weight functions for the
toroidal and vertical views, overlaid on a representative NUBEAM-computed fast
ion distribution function at $R=1.03$m are plotted in
Figure (\ref{fig:simfida}a). The dashed line indicates the
trapped/passing boundary. The toroidal views are more sensitive to
passing particles while the vertical views are more sensitive to the
trapped particles, however it is evident that at higher energies,
there are no trapped particles. Most of the signal from the vertical views at
higher Doppler shifts $\Delta \lambda$ therefore arises due to particles near
the trapped/passing boundary. This fact is particularly important when
considering the results presented in Section (\ref{sec:vertdata}), which
indicate that there is a deficit in the vertical signal at higher
Doppler shifts but not in the horizontal signal.  It is also
worth noting that the weight function is only non-zero for
$E>E_{\rm min}$ as described above, and that this shifts up or down
depending on the Doppler shift $\Delta \lambda$.  For these views, the
simulated FIDA signals are shown in Figure (\ref{fig:simfida}b).
Because of the co-injected beams, the toroidal FIDA signal is completely
red shifted. 

In addition to FIDA, there are beam emission components at full, half,
and one-third injection energy (from molecular deuterium in the
accelerator). The factor $\cos \theta_{lm}$, where $\theta_{lm}$ is
angle between the line of sight and the injected neutral beam $m$,
which determines the wavelength of the beam emission components, is
plotted in Figure (\ref{fig:dotp}b).  An optimum FIDA view would have
$\cos \theta_{lm}=0$ everywhere, so that the beam emission is at the
same wavelength as the background $\mathrm{D}_{\alpha}$ light, and
having a larger wavelength range containing fast ion information
\cite{heidbrinkorig}. This is the case for the SS vertical views near
the edge, but towards the core $\cos \theta_{lm}$ increases (because
the lens is mounted near the edge). Beam emission in the toroidal
views has a larger Doppler shift, with $\cos \theta_{lm}=0.8$ near the
innermost radius of 0.8m, thereby overlapping strongly over the FIDA
signal, however sufficient uncontaminated windows remain that a FIDA
signal may still be observed (see Section \ref{sec:measuredspectra}).

Finally, the factor $\cos \theta_{bm}$, where $\theta_{bm}$ is the
angle between the magnetic field and the injected neutrals, defines the
pitch parameter at birth of the fast ions. This is plotted as a
function of radius in Figure (\ref{fig:dotp}c). Comparison
with the $(R, p)$ resolved TRANSP-calculated fast ion distribution for energies above 30keV shows that the distribution is dominated by particles around the magnetic axis ($\sim 1.0$m); completely passing
particles are born near beam tangency (0.7-0.8m) inboard from the magnetic axis, and these follow trajectories outboard from the magnetic axis to produce a peak in the passing particle population at $\sim 1.2$m.  

\subsection{Spectrometer/CCD}

\begin{figure}[htb]
\includegraphics[width=8cm]{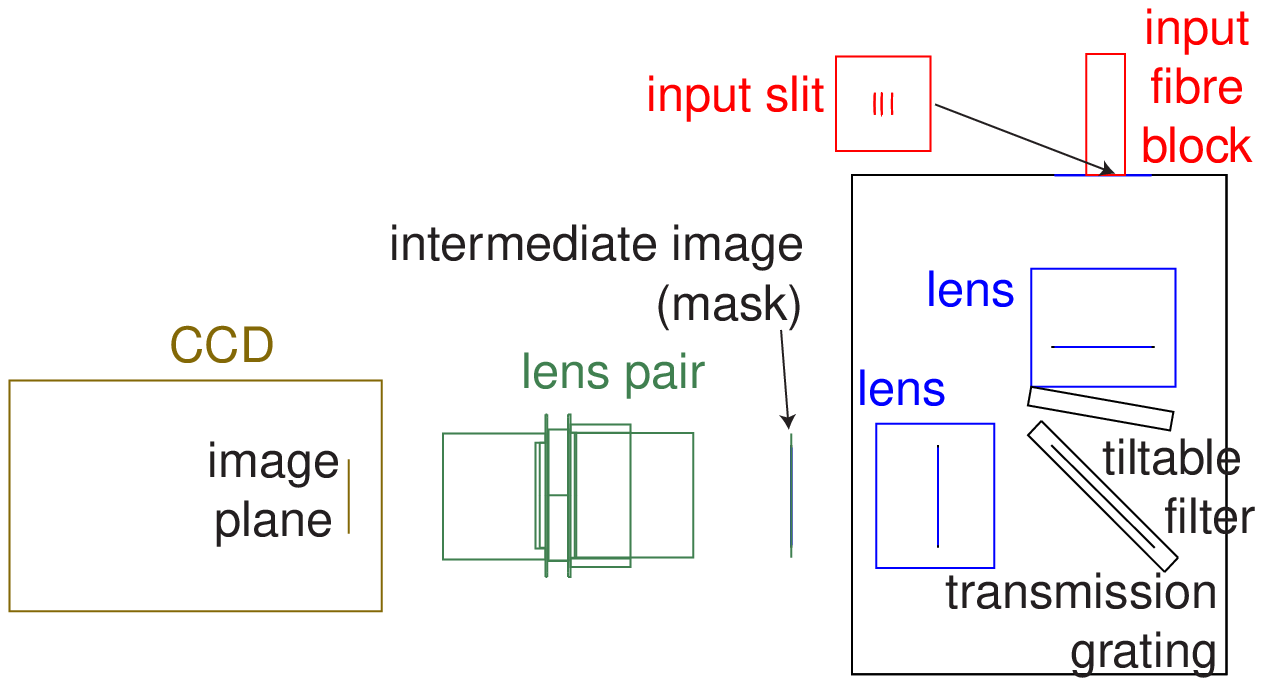}
\caption{Schematic of the spectrometer and CCD showing the fibre array, tiltable filter holder, grating, intermediate image location where masks can be inserted, second relay lens array and CCD.}
\label{fig:spectrometer}
\end{figure}
The spectrometer system resembles closely the system installed on NSTX
\cite{fidavertnstx}. The system installed on MAST is shown in
Figure (\ref{fig:spectrometer}). A Holospec f/1.8 f=85mm transmission
grating spectrometer \cite{bellholospec} is used which matches
optimally to the numerical aperture of the optical fibres. Instead of
a CCD at the output of the spectrometer, a second relay lens system is
incorporated which allows access to the intermediate image for
blocking unwanted bright spectral components. The second lens
system enables demagnification via a 85mm/50mm (Nikon f/1.8 / Nikon
f/1.2) lens pair. A conventional 2155 lines/mm grating is used (CWL:
653nm) which features a modest dispersion of 6nm/mm, giving a trade-off
between wavelength coverage and spectral resolution. An
EMCCD is utilized on the output, model Cascade 128+ from Photometrics.
This features a very fast readout rate of up to 3.3kHz (when binned
over 12 vertical channels), and vertical transfers of 80ns/row to
minimise frame transfer smearing.  As the efficiency of the system
is high, a typical FIDA spectral radiance of $10^{16}
\mathrm{ph/s/sr/m^2/nm}$ should theoretically give around 900
photo-electrons per pixel per 0.3ms time interval, which, with the use
of EM gain to mitigate read noise, should give a SNR of around
30. Additional losses have however been discovered, as outlined in Section \ref{sec:caltestsrel}.  One disadvantage of this CCD is that the chip size is somewhat small
($3 \times 3\mathrm{mm}$) compared with other instruments based on the
same spectrometer (e.g. \cite{fidavertnstx}, which utilizes an $8
\times 8\mathrm{mm}$ chip). The disadvantage of this is that fewer
channels can be accommodated (2 columns of 12 fibres), and
wavelength range is reduced to 20nm total for both columns versus about 50nm for the 8mm chip. According to Figure (\ref{fig:simfida}b), the minimum
required spectral band is approximately 650nm-662nm plus extra for
measuring the bremsstrahlung background. As the dimensions of the
chip are insufficient to accommodate the required wavelength range, it
was decided to use filters with reduced spectral bands of 6nm FWHM, spanning
649nm-655nm for the vertical channels (ignoring red-shifted
components) and 657-663nm for the toroidal channels.  As such,
blue-shifted vertical and red-shifted horizontal views cannot be
simultaneously measured. As this is desirable, a third column of
fibres was put in, and a filter used to pass the entire
14nm FWHM region from 649-663nm (including the unshifted
$\mathrm{D}_{\alpha}$ peak). The input focal plane consists of a
block with three columns of 12 fibres, on a curved radius to cancel
the vertical wavelength dispersion. The intention is that only two
columns are used at any one time, the closer pair being used for the
6nm FWHM bandpass filters and the further-apart pair used for the 14nm
FHWM filter. A translation stage permits the CCD and final image plane
to be translated with respect to the intermediate image plane to
adjust the wavelength region from the red filter band to the blue
filter band. Fixed, blackened slits of widths 50, 100 and 200$\mu m$
are available, giving spectral resolutions of 0.3, 0.6 and 1.2nm respectively
(however lens aberrations create additional `wings' to the instrument
function). The narrowest option is used for the toroidal views where
the beam emission contaminates the FIDA, while wider options are used
for the vertical views.  The filter is mounted inside the spectrometer
between the first lens and the grating, via a tiltable filter holder.
Because the unshifted $\mathrm{D}_{\alpha}$ light is known to be
several orders of magnitude brighter than the FIDA emission, it is
imperative to make efforts to avoid contamination of the FIDA signal.
Accordingly, the red and blue filters have $1\%$ transmisson at
656.1nm \cite{barr}, increasing to $>33\%$ 0.7nm away (with peak
transmission $~90\%$).  After the filter, the unshifted
$\mathrm{D}_{\alpha}$ is measured to be only about as bright as the
beam emission components.  With such a bright line in the wing of the
filter, back reflections off the filter may reflect back off the slit
mask area and pass though the filter at a steeper angle, where the
blue shift arising due to the tilt may cause the light to pass though
producing a contaminant line. The filters were therefore inserted at
an angle of a few degrees to the optical axis in order that the back reflection
effectively be off the side of the chip, and that the back reflection
hits the blackened area of the slit and avoids the other fibre columns.

The next problem that can arise is due to frame transfer smearing.
Although this is small (as the total frame shift time into the readout
area is only $10\mu\mathrm{s}$), the beam emission peaks in the
toroidal channels may smear into neighboring channels which contain
only FIDA and are naturally free from other contaminant lines.  This
is often a problem since the beam emission is about 2
orders of magnitude brighter than the FIDA emission, giving a smeared
signal comparable to, but somewhat less than, the FIDA emission.  To
mitigate this, a mask can be inserted at the intermediate
image plane with 'blocking bars' for the beam emission components.
Alternatives include the use of a FLC shutter as well as longer
integration times. In this respect, utilizing a smaller CCD gives
shorter row transfer times, reducing smearing when the integration time
is not set by the maximum rate of the camera.

\subsection{Patching layout}

While the diagnostic can only accommodate 24 fibres simultaneously,
there are 160 fibres available to be patched.  Given that at any one time the system is only imaging either toroidal channels or
vertical channels, there are available 32 active and 32 reference
fibres for the 24 spectrometer channels, each one with a spatial
resolution of $\sim 1.5$cm. Therefore, generally, every 3rd fibre is
patched in the system. Because of the finite lifetime of the excited
state, the intrinsic spatial localization of the
emission is around 2cm (though with broader tails). The Larmor
radius for the fastest ions is $\sim 5cm$ which limits the minimum
possible scale length of the fast ion distribution.
Patching neighboring channels would not therefore allow sharper gradients to be measured. Because the most complete information about the fast ion distribution is obtained when data are available from both horizontal
and vertical views, it is necessary to run identical discharges with
the system using the red and blue shifted filters on alternate shots.
This was achieved for a limited set of shots during the M8 campaign.

\subsection{Measured spectra}
\label{sec:measuredspectra}
\begin{figure}[htb]
\includegraphics[width=8cm]{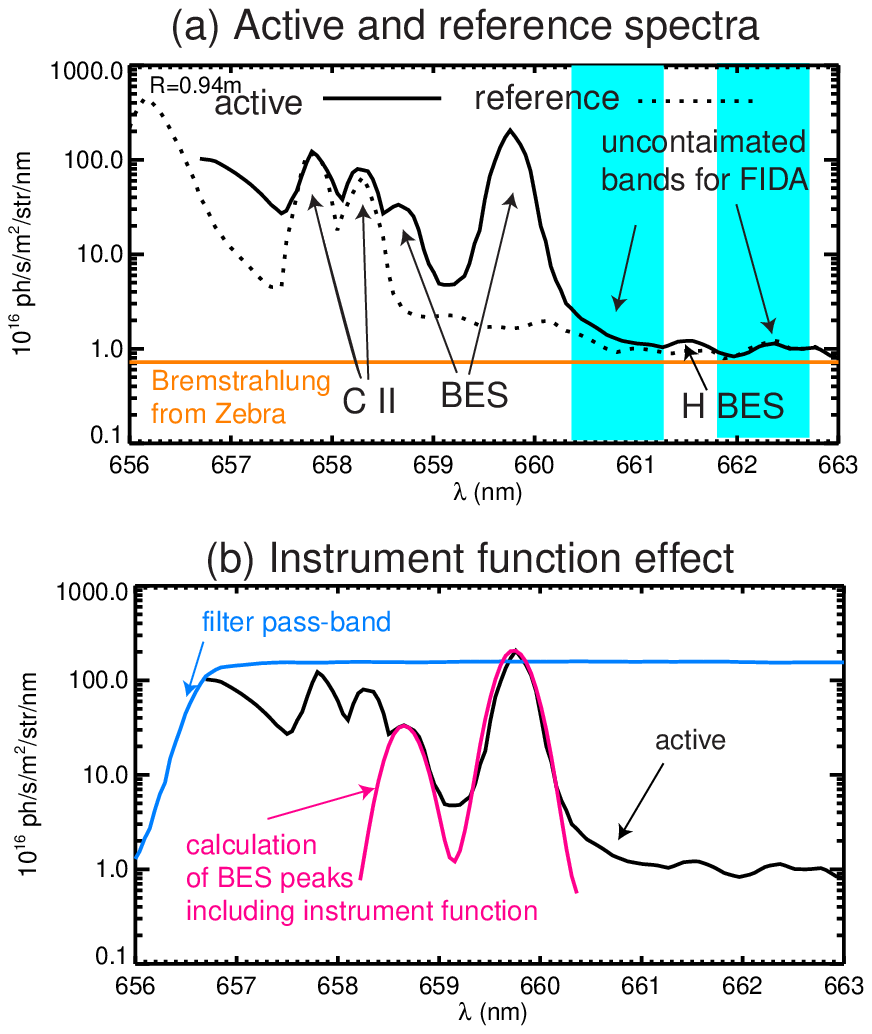}
\caption{(a) Active and (toroidally-displaced) reference spectra at
  $R=0.94$m and computed bremsstrahlung background (scaled from Zebra
  diagnostic, considering wavelength difference), highlighting the
  uncontaminated spectral region available for FIDA: between the BES
  peak and the H BES peak; and above the H BES peak, which does not
  contain any FIDA at this radius, but does contain FIDA in edge
  channels.  (b) Active spectra, compared with the synthesized BES
  peaks considering the measured instrument function width, as well as
  the filter pass-band.}
\label{fig:fidator1}
\end{figure}

Measured active and reference spectra for an inner channel of the
toroidal view are plotted in Figure (\ref{fig:fidator1}). The spectra
have been calibrated as described in the next section. The notable
features of the active spectrum include the BES peaks, including a
small amount at $\sqrt{2}$ times the primary Doppler
shift from hydrogen in the beam, C II impurity lines at 657.7nm and 658.3nm, and the unshifted $\mathrm{D}_{\alpha}$ component at 656.1nm. Both spectra are elevated by an amount very similar to that expected from bremsstrahlung considering $Z_{\rm eff}$ from the Zebra diagnostic and the difference of the measured wavelength region ($\sim 656$nm instead of $\sim 530$nm) \cite{zebra}.  

The differences between the active and reference spectra include the
FIDA emission, which should have a spectrum as shown in
Figure (\ref{fig:simfida}b), as well as the primary and hydrogen beam
emission spectrum (BES). The brightest contaminant is the unshifted
$\mathrm{D}_{\alpha}$. The filter bandpass, over-plotted in (b),
indicates how much the edge $\mathrm{D}_{\alpha}$ and lower energy
halo components are attenuated. The FIDA radiance is of order
$10^{16} \mathrm{ph/s/nm/m^2/sr}$. The edge $\mathrm{D}_{\alpha}$,
which should be about 10,000 times as bright as the FIDA signal, is
attenuated by the filter such that it is comparable in magnitude to
the beam emission peaks. The carbon impurity lines are much brighter
than the FIDA emission. Consequently, small errors in the background
subtraction lead to large errors in the FIDA such that these wavelengths
must be rejected. One interesting feature is that the passive spectrum is
not flat as one would expect if it were dominated only by impurity
lines and bremsstrahlung; rather, it has a spectral shape in the
region of 660nm which greatly resembles that of the active FIDA signal
as shown in Figure (\ref{fig:simfida}b). This wing indicates the
presence of passive FIDA driven by charge exchange between fast ions
and neutrals penetrating from the edge. The magnitude can be as large
as the active FIDA signal, and is examined in more detail in
Section (\ref{sec:subtraction}).

In Figure (\ref{fig:fidator1}b), the active spectrum is compared with the synthesized beam emission spectrum considering the measured instrument function width. It is clear that the width of the BES peaks is completely dominated by the instrument function. When plotted on a log scale, the spectral width
of the beam emission is approximately 0.75nm at the $1\%$ width level, while the theoretical instrument resolution is only 0.2nm. The reason
for the larger width is the presence of 'wings' in the point spread
function of the imaging lenses, particularly given that they operate
at a low f-number (f/1.2). Because the beam emission is approximately 50
times brighter than the FIDA + background emission, this eliminates
approximately 0.8nm of bandwidth from the usable spectrum. The signal
between the primary and secondary beam emission peaks is a little
higher than the predicted `wings', which may be due to FIDA emission,
but it is very difficult to isolate. Thus only wavelengths higher than the beam emission peak are taken.  

The H BES peak is very weak compared with the main BES peak, and can
vary depending on the the degree of getter pumping in the NBI. It can
introduce significant errors when the FIDA signal is calculated.
Reasonable FIDA data estimates can however be obtained by interpolating over the H BES peak; this was done for later figures showing the radial
profiles of the net FIDA signal (active minus reference).  

Spatial profiles of the net FIDA signal are
plotted in Figure (\ref{fig:fidatordnd}c) for different wavelengths
corresponding to $E_{\min}$ from 30 to 60keV. In this figure, regions
influenced by beam emission have been removed (no lines or points),
and those influenced by H beam emission have been interpolated over
(lines, but no points). Because the Doppler shift of the beam
emission varies with channel, lower wavelengths
are absent for inner radii.

\begin{figure}[htb]
\includegraphics[width=8cm]{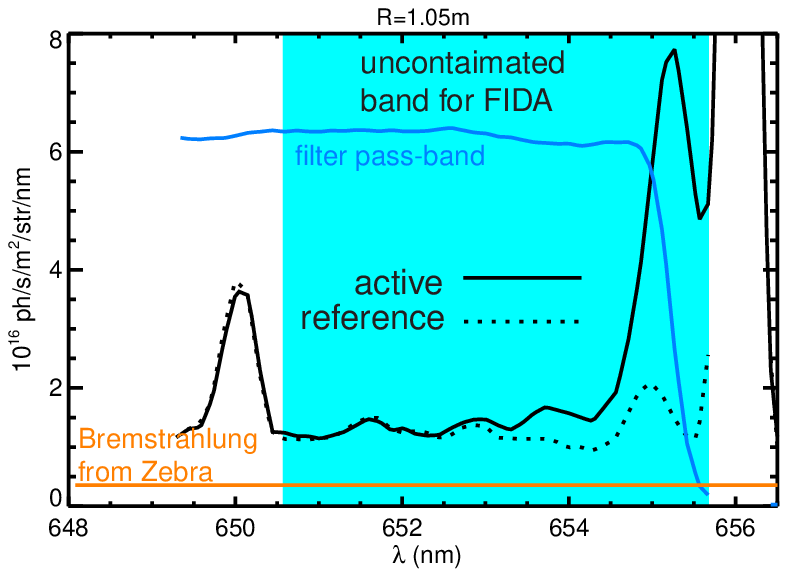}
\caption{Measured active and reference spectra for an inner channel of the vertical-viewing fibres, compared with the filter pass-band and the calculated bremsstrahlung background.}
\label{fig:fidavert1}
\end{figure}

Active and reference spectra from the vertical view are shown in
Figure (\ref{fig:fidavert1}), compared with the expected bremsstrahlung
level based on Zebra measurements, and the filter pass-band. There is a peak at about 655.2nm, corresponding to
a competition between increasing halo closer to the unshifted line,
and decreasing filter transmission.  The available wavelength band for
the vertical views is much larger, the entire blue shifted region
being free from contaminants apart from an oxygen line at 650nm which
is out of the FIDA spectral region. The background is however
somewhat larger than the bremsstrahlung level, probably due to
radiation from the divertor region.

\section{Calibration and tests}

\subsection{Relative calibration}
\label{sec:caltestsrel}

A relative calibration was made between the tokamak fibre and coupling
lens system and the calibration fibres located adjacent to the
spectrometer.  Separate calibration factors were obtained for each
individual fibre and were found to be approximately $60\%$, with the
channel-to-channel variation being within $\sim 5\%$ and additional
reproducibility errors being of order $10\%$. The windows are shuttered
during discharge cleaning and have a transmission of $\sim 95\%$.

Calibration of the vertical views is much more difficult than the
horizontal views on account of the fact that (1) the internal lenses are
not protected from deposition from the plasma and glow discharge
cleaning pulses, and (2) the vacuum feed through system relies on
imaging fibre to fibre and is prone to small misalignments. The
coupling efficiency of the in-vessel lens to the ex-vessel fibres is
$\sim 13\%$, as determined by the ratio of the beam emission compared
with that expected from the beam model, while it was measured to be
$\sim 30\%$ during installation. The reason for this discrepancy may be due to coating of the in-vessel lens, however the transmission has not
deteriorated significantly during an experimental campaign.

For each viewing bundle, active to reference cross-calibration can be
performed by looking at discrete spectral lines as well as the
broadband bremsstrahlung baseline. For the toroidal views, both broadband and line components are very similar between active and
reference views, indicating that no additional calibration factor is
necessary. The vertical active signals require on average a relatively large
scaling factor of 2.5 to bring them into agreement with the vertical
reference views. This is similar to the observed reduction in
transmission efficiency of the active lens ($30\%/13\%$), however this ratio can vary substantially between indivdual pairs of (active/reference) channels and may be indicative of the spatial structure of divertor molecular emission.

To account for this relative calibration, routine analysis
presented in later sections includes multiplication by the basic
active/reference transmission ratio (which for the toroidal view is
close to unity) and offsetting of the reference emission to force the
background level to be the same (at 662nm) in active and reference views, in order that basic mis-matches (due to reflection of broadband molecular emission from the divertor, for example) are avoided.

\subsection{Absolute calibration via beam emission}
\label{sec:beamemission}

\begin{figure}[htb]
\includegraphics[width=8cm]{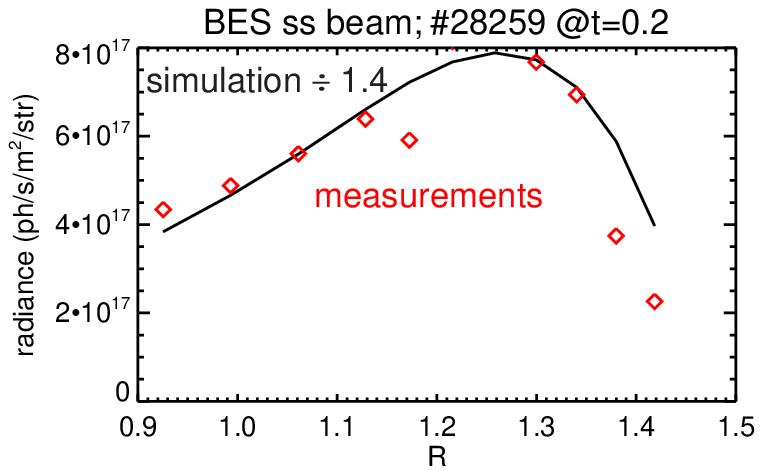}
\caption{Measurements (points, red) and modelling using latest ADAS coefficients and the NEBULA code (line, black, scaled down by 1.4), of the radial profile of the primary beam emission peak as measured from the FIDA spectrometer.}
\label{fig:bes}
\end{figure}

Because the FIDA spectrometer is designed to view the weak FIDA light,
the instrument must be configured deliberately to obtain beam
emission measurements on the toroidal views without saturation of the detector. The profile of the simulated (using the NEBULA code, as described in Section (\ref{sec:calc})) and measured beam emission is compared in Figure (\ref{fig:bes}). Considering the relative calibration factors above,
and absolute calibration of the spectrometer, a scaling factor of 1.4
is required to model the beam emission. On the other hand, the
broadband bremsstrahlung component often agrees with that calculated
from the Zebra diagnostic (partially as indicated in
Figure (\ref{fig:fidator1}), though other analysis has confirmed this
more thoroughly), indicating consistency of the absolute calibration.
The spatial profile of the beam emission matches well with the model,
indicating that the beam attenuation is modelled correctly, which
validates the spatial profile of the power and fast ion deposition.
It is evident that the disagreement factor of 1.4 is indicative that
at least one of the assumptions made above is wrong, those being:
absolute calibration of the FIDA diagnostic; divergence profile of the
beam, beam power, electron temperature and density; FIDA sightline
geometry with respect to the beam; and atomic physics rate
coefficients.

The beam specification is given in \cite{gee}. The shape (divergence) and position of the beams are confirmed with an unfiltered camera which can clearly see the beam during beam-into-gas discharges, and another filtered
imaging diagnostic which measures the Doppler
shifted beam emission in the range of 660nm. Beam power
measurements take account of a neutralization model for the beam
and measurements of the species fraction mix. Modelling of the beam
deposition is likely to be accurate to better than 10 percent, and can be confirmed by the profile shape of
beam emission as seen in Figure (\ref{fig:bes}).

Geometry of the sight lines with respect to the beam was confirmed
using backlighting tests and indicate that the alignment is correct to within
about 2cm, which should degrade the beam emission by only $5\%$. The
beam power is considered to be correct on two accounts; firstly since
the measured total neutron rate agrees with TRANSP simulations in MHD
quiescent discharges, and secondly, since the toroidal FIDA signals
(scaled to match the beam emission) also agree with TRANSP simulations
in these same quiescent discharges (see for example
Figure (\ref{fig:fidator1})). This is because the BES signal $\propto
P$ while the FIDA signal $\propto P^2$ (where $P$ is the beam power).
A calibration error would affect these two signals by the same factor
but a power error would affect the FIDA signal more strongly.

The ADAS rate coefficients used here were shown during experiments on
JET to be discrepant by around the same factor of 1.4 derived here,
but in that work the discrepancy was attributed to uncertainty in the
absolute calibration. On the other hand, on DIII-D, the rate
coefficients have been confirmed \cite{grierson}. On ASDEX too, FIDA
measurements have been shown to be in absolute agreement with
modelling, indicating that the some of the rate coefficients are
correct (beam emission modelling is only based on electron and ion
collision cross-sections, while FIDA modelling relies on the same
cross-sections as well as state-selective charge exchange).

\subsection{Passive FIDA subtraction}
\label{sec:subtraction}
As shown in Figure (\ref{fig:fidator1}), there exists passive FIDA
emission even on the reference views caused by edge neutrals
\cite{heidbrinkbestoonstxd3d}. The amount of passive FIDA is
experimentally estimated to be the spectrum minus the minimum of the
spectrum, around 662nm, which is taken to be the (bremsstrahlung and molecular emission)
background. The amount of passive FIDA is in delicate balance,
influenced by the penetration depth of neutrals from the edge as well
as the population of fast ions towards the edge of the plasma.
Performing beam cut-off experiments and examining the on-off
difference is a method to independently check the amount of passive
emission in the active view and compare that with the reference views,
as immediately after beam-switch off, a substantial population of fast
ions remains. On the other hand, at the initial beam-switch on, there
is no passive FIDA component because of the lack of fast ions.

\begin{figure*}[htb]
\includegraphics[width=18cm]{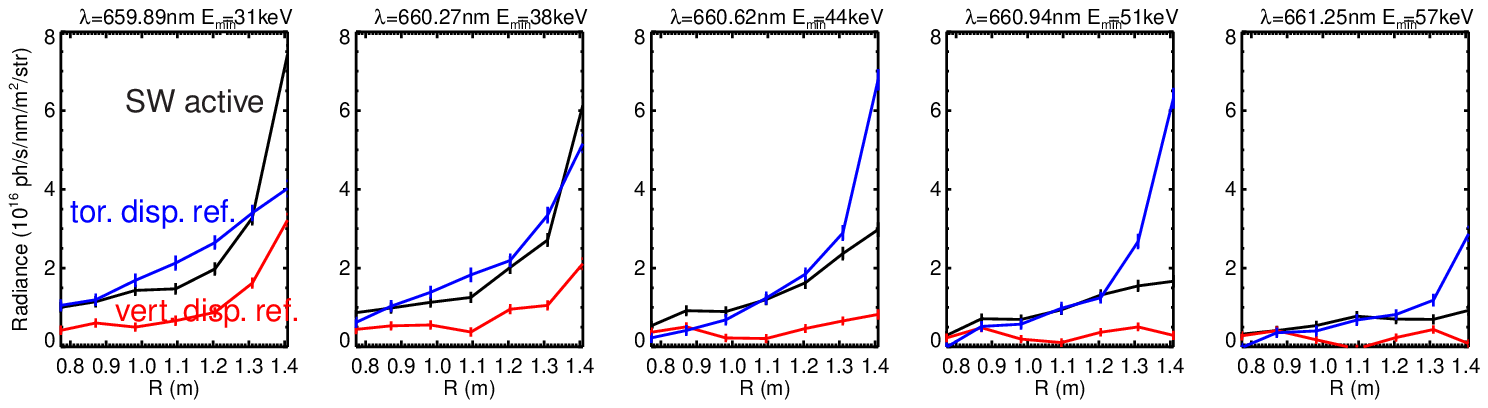}
\caption{Comparison of radial profiles of passive toroidal FIDA signals from the active view (installed on the upward-directed SW view) derived just after the beam switch off; from the toroidally-displaced, downwards-directed reference view; and from the vertically-displaced, upwards-looking reference view. Data are at $t=0.24s$ on $\#28141$ (for which two reference views were simultaneously acquired).}
\label{fig:pass1}
\end{figure*}

\begin{figure*}[htb]
\includegraphics[width=18cm]{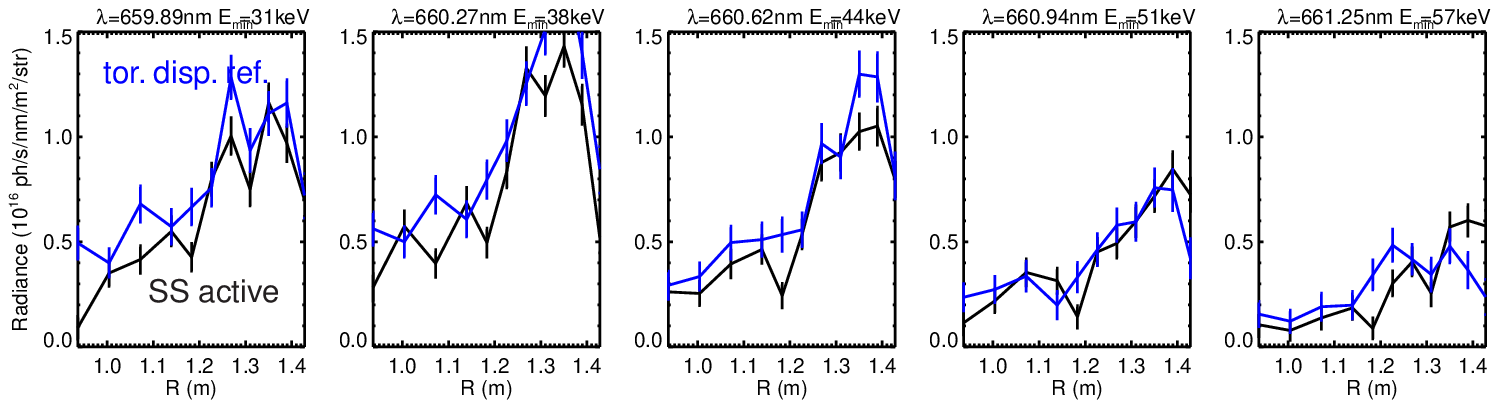}
\caption{Comparison of radial profiles of passive toroidal FIDA signals from the active view (installed on the downwards-directed SS view) derived just after the beam switch off and from the toroidally-displaced downwards-directed reference view, taken at $t=0.205$s on $\#28322$ (similar to chirping mode shot $\#28319$ shown in Figure. (\ref{fig:fidatordndfb})).}
\label{fig:pass2}
\end{figure*}

Radial profiles of the passive FIDA (calculated as above) on the
active and reference bundles, before and after beam switch off, and for
various wavelengths are plotted in Figures (\ref{fig:pass1}) and
(\ref{fig:pass2}). In Figure (\ref{fig:pass1}), the toroidal lens was viewing the SW beam. An alternative tokamak-spectrometer patching
scheme was used in which there were 7 chords each from the active bundle, vertically-displaced reference bundle and toroidally-displaced
reference bundle. It may be observed that there are noticeable
discrepancies between active and both vertically and toroidally-displaced reference bundles. The vertically-displaced reference view
has a very similar spectral shape to the active view but has a serious
deficit in the magnitude of the passive FIDA, which renders this view
most inaccurate for background subtraction since the magnitude of the
passive FIDA is similar to the active FIDA. This difference in
magnitude may be due to the vertical extent of the fast ion
distribution or the donor neutral density and motivates the
development of a model for passive FIDA emission.

The magnitude of the passive FIDA on the toroidally-displaced reference view is closer to that in the active view, but is slightly stronger towards the edge in the 660-661nm region. This extra passive emission may be caused by the vertical tilt of the toroidally displaced reference ($-6^{\circ}$), this being significantly different from that of the active view ($+8^{\circ}$).
Referring to Figure (\ref{fig:dotp}), $\cos \theta_{b l}$
of the the SS beam (which is tilted down by $-8^{\circ}$, similar to that of the toroidally-displaced reference) is close to the pitch parameter 
characteristic of the birth of fast ions, near the edge, of about 0.7 so it is
sampling parts of the fast ion distribution which have not
been scattered in pitch angle. The energetic parts are scattered only
by electrons which affect their energy rather than their pitch angle.
On the other hand, $\cos \theta_{lb}$ of the SW beam is
very close to unity and rather different from the birth pitch parameter, indicating that this picks up less of the signal from recently-born fast ions.  The passive emission from SW active views and toroidally-displaced passive might therefore plausibly be different.

A similar background subtraction test was conducted (in different
discharge conditions) between the active bundle on the SS beam and the
toroidally-displaced reference in Figure (\ref{fig:pass2}). Here, the
passive emission profile is very similar in both magnitude and shape
on the SW view. There are still some discrepancies at
higher wavelengths on the edge few chords. This
configuration was therefore chosen for further analysis in this paper.

\section{Forward modelling of FIDA and beam emission}
\label{sec:calc}
Thorough modelling of FIDA and beam emission can be done with the
FIDASIM code \cite{fidasim}. This is based on output from the TRANSP
code, which uses the NUBEAM module for the fast ion distribution
function. In its basic form, NUBEAM considers only classical collisional
slowing down and transport. Additional anomalous diffusion
$\mathrm{D}_{\rm an}$ coefficients can however be introduced
using a variety of parameterisations, which may be utilized to match
the experiment due to anomalous activity from fast
particle MHD, for example. There will certainly be a broader class
of models which would also be consistent with the results, but these
simple models are useful  for evaluation of the
fast ion heat source.

For TRANSP simulations, MSE-constrained EFIT equilibria are
used. Thomson scattering data give electron temperature and density;
CXRS provides ion temperatures and bulk toroidal rotation velocities;
and the Zebra bremsstrahlung diagnostic is used to obtain profiles of
$Z_{\rm eff}$. All diagnostics have spatial resolutions
better than a few centimetres at most. Profiles are mapped to the EFIT equilibrium and fitted on both sides of the magnetic axis using adaptive
smoothing based on error bars and goodness of fit. For rapid modelling of
beam emission measurements incorporating up-to-date ADAS rate
coefficients, the NEBULA code was used. This is based on the code NEMO
\cite{nemo}, formerly used for the same purpose \cite{ephrem}. This code has been validated against other codes, in particular FIDASIM and TRANSP.

\section{Comparison of FIDA measurements with modelling}

\subsection{Toroidal view in MHD-quiescent phase of a low-power discharge}

\begin{figure*}[htb]
\includegraphics{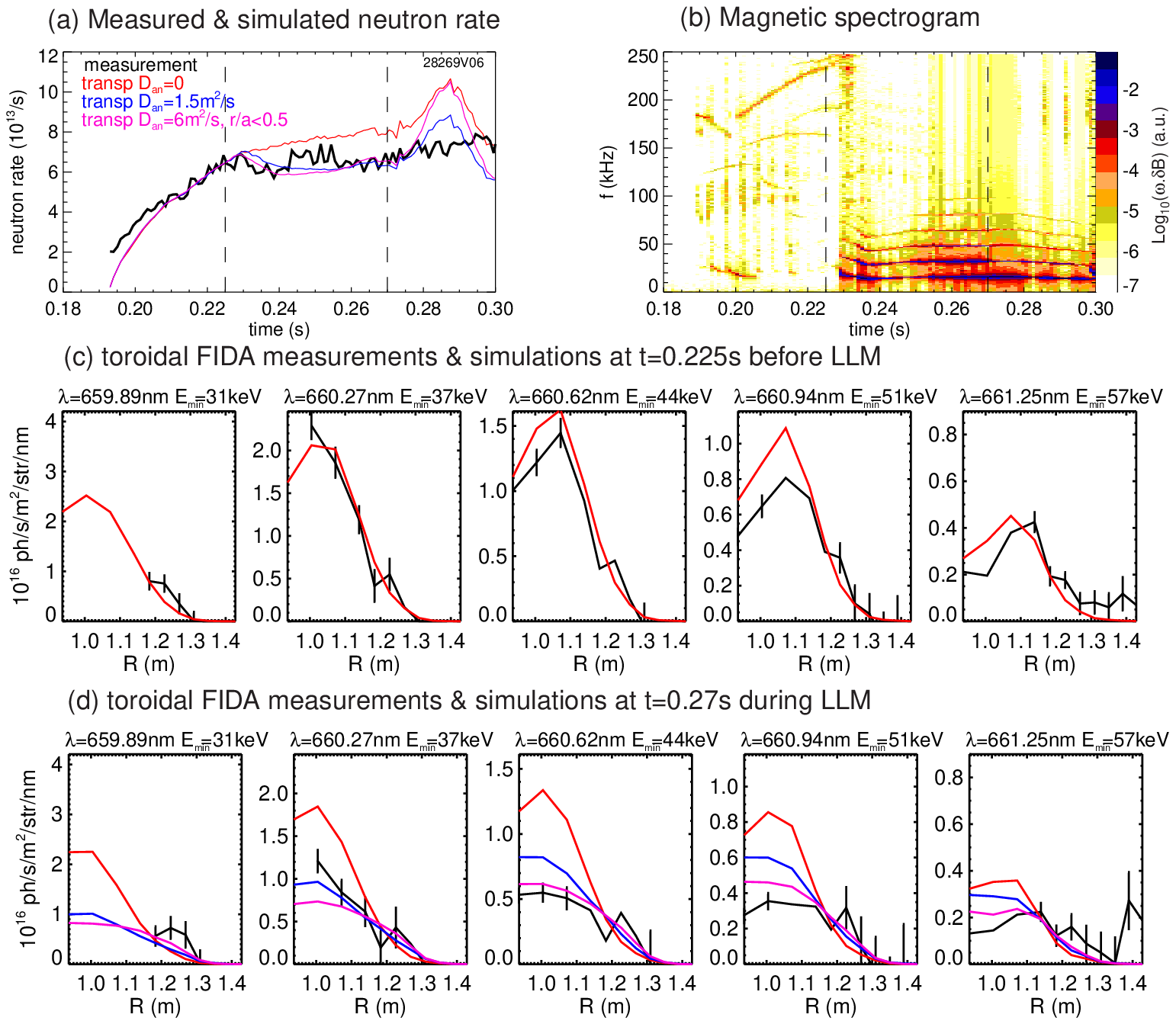}
\caption{Comparison of measured fast ion data with simulations in $\#28269$. (a) Time trace of neutron rate and simulation assuming various diffusion models. (b) Magnetic spectrogram showing the onset of a LLM at 0.23s. (c) Toroidal FIDA measurements at different wavelengths and classical modelling during the quiescent phase. (d) Comparison of measurements with modelling considering different diffusion models during the LLM (the colours corresponding to particular models are the same as those used in (a)).}
\label{fig:fidatordnd}
\end{figure*}

A period within a discharge is chosen in which the neutron rate matches
the TRANSP calculation. The time evolution of the neutron rate,
compared with TRANSP simulation, as well as a magnetic spectrogram is
plotted in Figure (\ref{fig:fidatordnd}a). This discharge features a
long lived mode (LLM) after $t=0.23$s, but in this section
$t=0.225$s (during the quiescent phase) is analysed. Spatial profiles
of the FIDA signals are shown in Figure (\ref{fig:fidatordnd}c) for
different wavelengths corresponding to $E_{\min}$ from 31 to 57keV.
The error bar in the figure is a combination of the shot-noise error
bar as well as an estimate of the (subtraction) systematic error,
taken to be $20\%$ of the value of the passive FIDA signal as
described in Section (\ref{sec:subtraction}). FIDASIM code results based
on classical NUBEAM calculations (scaled down by 1.4 in order to match
the beam emission) are compared with the experimental data points. Though the overall agreement is not perfect here, it is quite reasonable, and serves as a basis for the investigation of anomalous effects.
The slight over-estimation at $E_{\min}=31$keV may be due to
frame-transfer smearing of the peaks; this occurs despite the use of a
blocking mask. There is a discrepancy with edge channels at higher
energy as a consequence of the error in passive background
subtraction; it was shown in Figure (\ref{fig:pass2}) that the
reference view often sees a larger signal than the active view at FIDA
wavelengths as a consequence of the slight difference in viewing
geometry. There is an anomalously high value in the few edge channels
at $E_{\min}=57$keV, while the subtracted signal goes negative
(because the passive signal is larger) at around $E_{\min}=44$keV.
These discrepancies become larger in the presence of MHD.

\subsection{Vertical system measurements in quiescent discharge}
\label{sec:vertdata}
\begin{figure*}[htb]
\includegraphics{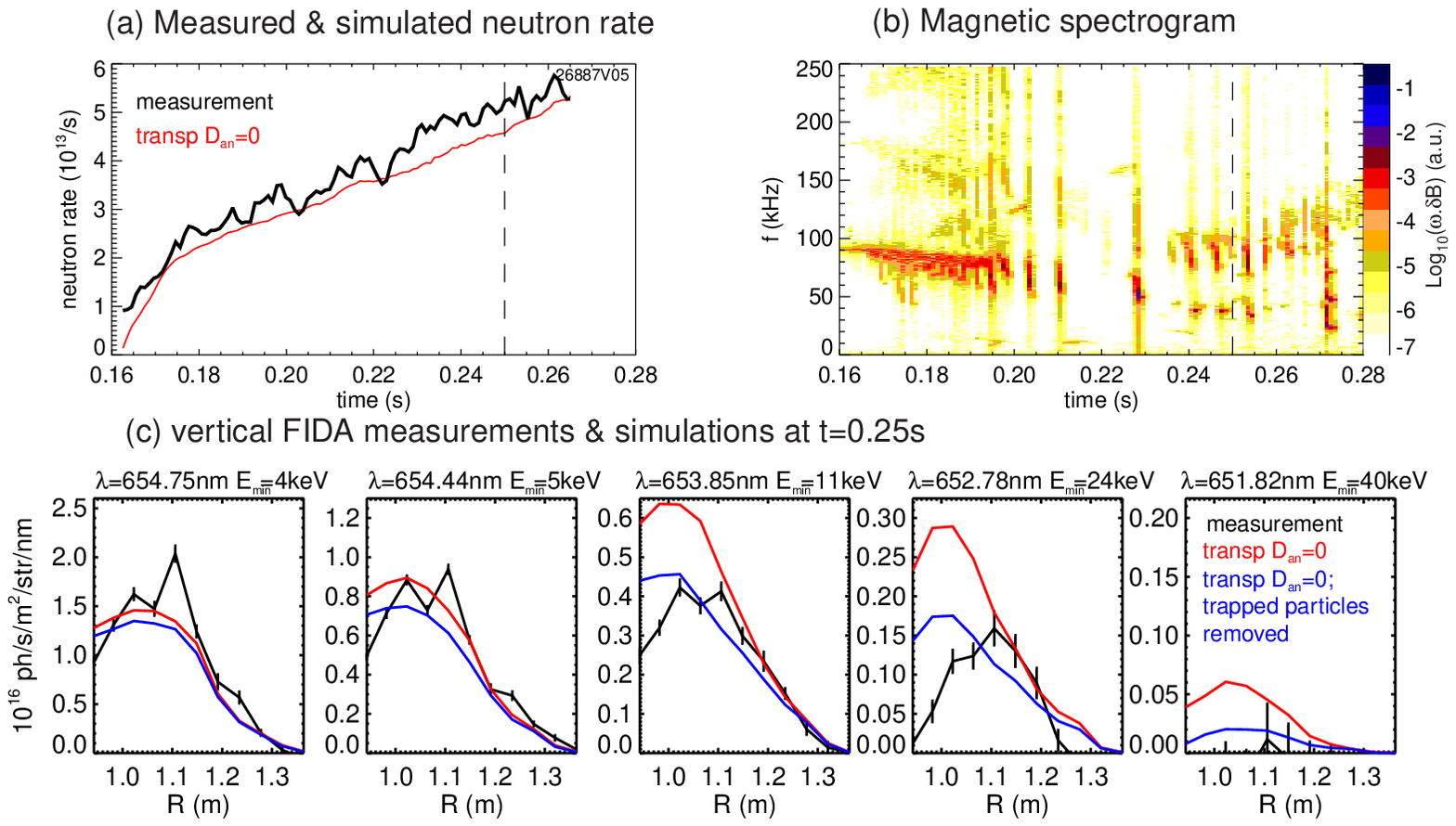}
\caption{Comparison of measured fast ion data with simulations in $\#26887$. (a) Time history of neutron rate and classical simulation. (b) Magnetic spectrogram showing weak chirping activity. (c) Vertical FIDA measurements at different wavelengths (much lower energies than the toroidal ones) and modelling assuming classical transport and also considering an absence of trapped particles, to elucidate their contribution to the signal.}
\label{fig:fidaverdnd}
\end{figure*}

\begin{figure}[htb]
\includegraphics{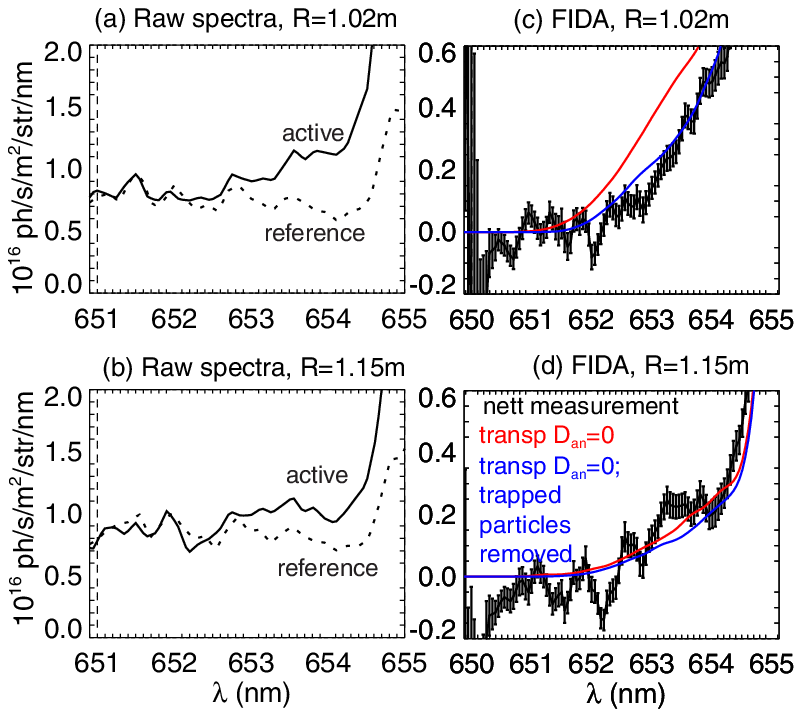}
\caption{Spectra at (a) inner and (b) outer radii on vertical channels for $\#26887$, and (c, d) comparison of the net active FIDA signal with simulations.}
\label{fig:fidaverdndspectra}
\end{figure}

The MHD-quiescent discharge $\#26887$ was part of a series during
which toroidal and vertical data were taken on repeat discharges,
however since the vertically-displaced reference was used for the
toroidal data, such data cannot be used. The use of both views
simultaneously would allow tomographic reconstructions of the fast ion
distribution to be made to complement the unknown parts of the
distribution function from each measurement \cite{mirko}. The time
history of the neutron rate and magnetic spectrogram is shown in
Figure (\ref{fig:fidaverdnd}a).  While the discharge does feature some
chirping modes, the neutron rate matches closely the TRANSP prediction
(the slight under-prediction may be a result of the fact that the
simulation start time is slightly after ($<1$ms) the actual beam start
time due to data timing issues). The FIDA signals are plotted in
Figure (\ref{fig:fidaverdnd}b). An attenuation factor of 1.4, derived
from the toroidal beam emission discrepancy, is used in addition to
value of $13\%$ due to the in-vessel fibres (n.b. this factor was
verified via beam-into-gas shots on the vertical view). Additionally,
the background has been multiplied by 0.4 to account for the coating
of the active in-vessel lens. The chosen values of $E_{\min}$ here are
much lower, simply by virtue of the fact that the higher values
contain little or no signal. The lowest value here is just in the tail
of the halo. Here, the proper halo calculation was
performed using the direct charge exchange module of the FIDASIM
code. There are noticeable discrepancies between the measurement and
the (classical) theory. It appears that the signal is discrepant by up
to a factor of 2 for $E_{\min} <~ 25keV$. For higher energies the
discrepancy is much larger, with there being no clear difference
between active and reference signals while theory would predict that
there should be. It is worth noting that the discrepancy does not vary
between the fishbone events, and is present in many other discharges
with a low amount of MHD activity where the neutron rate is classical.

The discrepancy in energy space is more clearly visualised by
examining the spectra in Figure (\ref{fig:fidaverdndspectra})
of raw signals (a), subtracted signals and simulations (b) for the
inner radii where the discrepancy is largest and a mid-radius channel
where the discrepancy is small. It can be seen that the spectral
shape at $R=1.02$m does not match the simulation, and that even a
scaling factor would not bring this into agreement. On the other
hand, at $R=1.15$m, the agreement is reasonable. It is noteworthy
that in both cases, the FIDA signal is somewhat smaller than the
background level.  

Because of the fact that trapped ion orbits have worse confinement properties than passing ion orbits, it is worth considering whether
this discrepancy might be explained by a reduction in trapped
particle fraction. To test this, the trapped particles were 'zeroed' out in
the TRANSP calculation result and fed into the FIDASIM calculation.
It is unlikely that this corresponds to the real situation, but it serves to demonstrate the contribution of trapped particles to
the FIDA signal. The result of this calculation brings the magnitude
of the signals into considerably better agreement, but there are still discrepancies particularly in the core region at $E_{\min}=24$keV,
equivalently for $\lambda < 652.8$nm. Note that reducing the trapped particle fraction results in a negligible reduction
of the predicted neutron rate as the highest-energy particles tend to be
passing.

Redistribution of the passing particles may also
reduce the FIDA signal. The fast ion distribution,
together with the vertical weight function for $E_{\min}=38keV$, is
plotted in Figure (\ref{fig:simfida}a) at $R=1.02$m, corresponding to
that in Figure (\ref{fig:fidaverdnd}b, far right panel). It can be seen
that the given wavelength for $E_{\min}=38keV$ has its response
function just intersecting the primary birth energy peak at
$E=60keV$, $p=0.7$ (corresponding approximately to the birth pitch
angle). One way in which the simulated signal could be reduced is if
the average pitch parameter were higher, thereby `missing' more of the weight function of the vertical chords. Initially, it would appear
that such an explanation is unlikely, as the volume of phase space contributing to the weight function 
is smaller at larger pitch parameter. On the other hand, if one
inspects Figure (\ref{fig:dotp}c), at inner radii, close to the
tangency radius of 0.7m, the birth pitch parameter becomes close to
unity. The fast ions follow a flux surface to give a large component of the distribution at pitch parameter unity near $R=1.2$m. If the
fast ions were not 'born' until they were further into the plasma,
then their value of pitch parameter would be closer to unity.
Another possibility is that finite Larmor radius effects,
included only via a coarse approximation in the presented simulations, may play a role. This would particularly
affect potato orbits near the core, resulting in slightly different pitch
parameters. 

Other FIDA systems on other tokamaks have had differing levels of
agreement between modelling and data on vertical views. In DIII-D for example, agreement was
found with theory on vertical views (or horizontal ones sensitive to
the trapped ions) \cite{heidbrinkweight}, while on NSTX, there have
been problems gaining agreement on the vertical views, although calibration
errors contributed to this discrepancy \cite{heidbrinkhtpd}.

\subsection{Power scaling of vertical FIDA signals}
\begin{figure}[tb]
\includegraphics{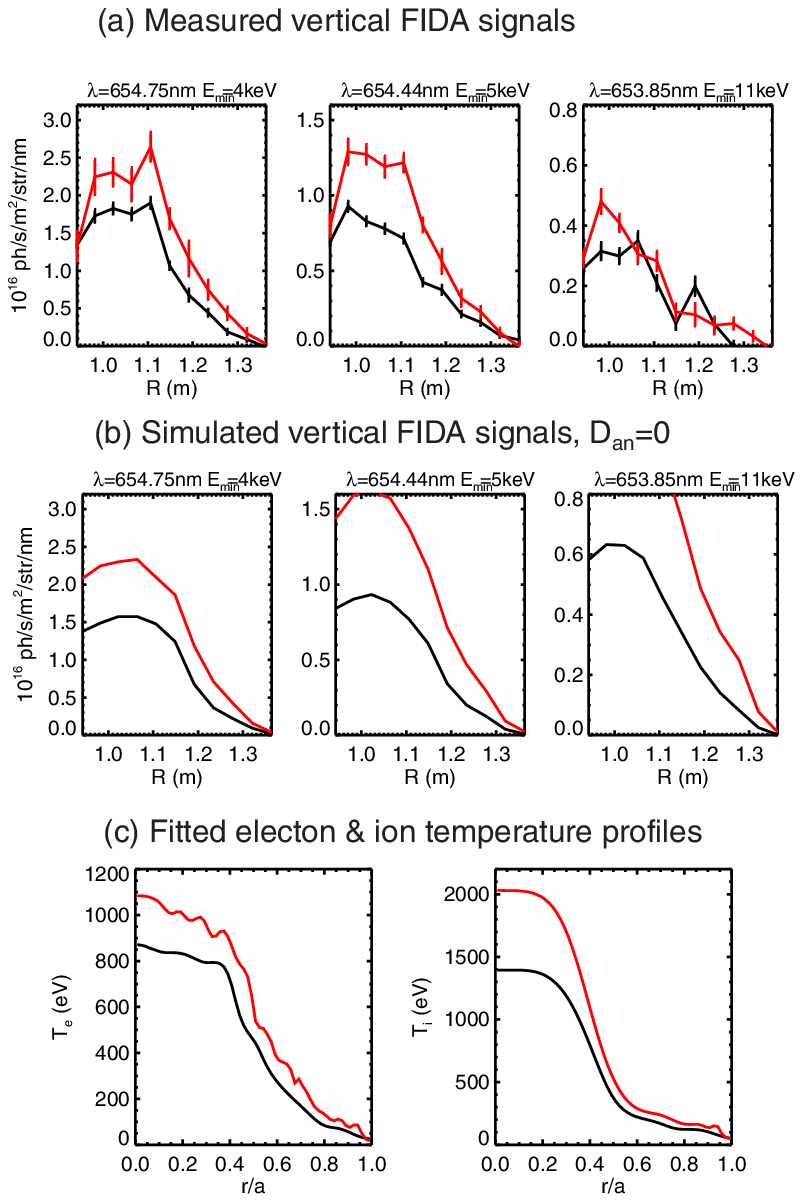}
\caption{(a) Comparison of low energy vertical FIDA signals between 1 (black) and 2 (red) beam shots $\#26887$ (where neutron rate agrees with modelling assuming $\mathrm{D}_{\rm an}=0$) and $\#26864$ (neutron rate matched by $\mathrm{D}_{\rm an}=1.5m^2/s$). (b) Simulated FIDA profiles. (c) Electron and ion temperature profiles input to TRANSP}
\label{fig:tempcompare}
\end{figure}

In the same set of experiments as the previous section, the discharge
was repeated with double the beam power, achieved by simultaneous
injection of the SS and SW beams. Here, the neutron rate goes up by a factor of 1.7. The comparison of the measured and (classical) simulated
vertical FIDA signals at $E_{\min}=4, 6, 11$keV, and ion and
electron temperature profiles, are shown in
Figure (\ref{fig:tempcompare}). The measured increase in vertical FIDA signals is approximately $50\%$ of the expected increase at 6 and
4 keV, while there is practically no increase at 11 keV and above.
Toroidal FIDA signals on the other hand do increase by a
reasonable factor when the beam power doubles.

The ion temperature increases by $\approx 40\%$ between the 1 beam and
2 beam shot, while the electron temperature increases only
slightly. As the ion heating rate scales as $E^{-3/2}$, such low
energy fast ions contribute to most of the ion heating. This is
evidence that the apparent stiffness in the bulk heat transport causes
or is caused by the same anomalous effects that degrade the confinement
of fast ions at energies only few times the thermal energy.

\subsection{Degradation of toroidal signals due to long lived mode}

In most discharges, a long lived mode (LLM) appears as q approaches unity \cite{iansat}. This is associated with strong braking
and a steady mode structure. While the LLM might be avoided in
advanced tokamak scenarios in future devices, it does indeed show how
the fast ion signal decreases in response to a symmetry breaking and
how certain resonant particles are lost. It is somewhat relevant to
the chirping modes in MAST which are often decreasing to the same
frequency as the LLM, and are shown theoretically to have a similar n=1
kink structure (though there are important differences). The measured
signal at $t=0.27s$ in shot $\#28269$ is plotted in
Figure (\ref{fig:fidatordnd}d). The measured neutron rate is $80\%$ of
the calculated neutron rate assuming $\mathrm{D}_{\rm an}=0$. On the
other, hand, FIDA measurements are discrepant by about $50\%$ in the
core region, and within error bars agree towards the edge except at
high energies where there is possibly a contribution from the incorrect
background subtraction discussed previously.  The y-scales in each
subplot corresponding to different values of $E_{\min}$ are scaled in
proportion to the contribution to the neutron rate. To assess the
effect of other models two further simulations were performed, one in
which $\mathrm{D}_{\rm an}=1.5m^2/s$ over the entire radius (for
$t>0.23s$), and one for which $\mathrm{D}_{\rm an}=6m^2/s$ for
$r/a<0.5$ (again for $t>0.23s$).  Both of these models match the
neutron rate. However, by using the model for which $\mathrm{D}_{\rm an}$ is applied only in the inner half radius, the modeled higher energy FIDA
signals match better to the measurement. On the other hand, at the
lower energy band wavelength, $E_{\min}=31kV$, this model has
decreased the signal by too great a factor.  Some sort of hybrid model
would be appropriate where the diffusion coefficient decreases at
lower energies. An energy dependent model has been used in the past
\cite{anthonytransport}, and tends to give a better match to the stored
energy while maintaining a similar neutron rate. Stored energy is not
compared here, as it is often found that in discharges for which the
neutron rate matches a classical TRANSP simulation the stored energy
is underestimated.

Such ad-hoc methods for applying a diffusion operator in TRANSP do not
completely elucidate the physics, as there may be other models which
replicate the same experimental signals. Nevertheless, simple
models give an idea of the likely heating power delivered
to the ions and electrons. For example, in this case, the heating
power at mid radius is only $5\%$ different between the two anomalous
models, validating the existing transport simulations at half
radius.

\subsection{Degradation of toroidal signals due to chirping modes}

Most discharges in MAST feature some sort of chirping modes, and they
tend to extend from the TAE range of frequencies down to the rotation
frequency of the core, corresponding to an n=1 kink mode
characteristic of a fishbone oscillation. The avoidance of these
modes is a current research topic and necessary for the design of
future high performance devices. On the other hand, for transport
analysis of the highest power discharges for example,
some chirping modes must be taken into account in the calculation of
the ion and electron heating power. While only the neutron measurements have been used for calculating the effective diffusion
coefficient (including the fission chamber and neutron camera \cite{marconc,mikhailiaea}), a range of functional forms for the spatial/energy
dependence of the diffusion coefficient may exist. In this respect the
FIDA data may corroborate/invalidate some of these models.

With any bursting, intermittent mode such as a fishbone or
sawtooth, the period between events is often larger when the events
are larger, so the net transport level may not depend strongly
on the period. It is important however to understand the trigger for
the mode as well as the mechanism for the degradation of fast ions.
Fishbones are considered to be triggered by trapped ions but can
redistribute a significant fraction of passing ions. This is determined from the fact that, for example, the neutron rate is affected, which is known to be most sensitive to the higher energy passing ions. The relaxation between
fishbones is therefore likely to be governed by the 
collisional phase-space redistribution of particles back towards equilibrium.
Often however, the magnetic spectrogram at low frequencies does not go completely quiet between fishbones.

\begin{figure*}[htb]
\includegraphics{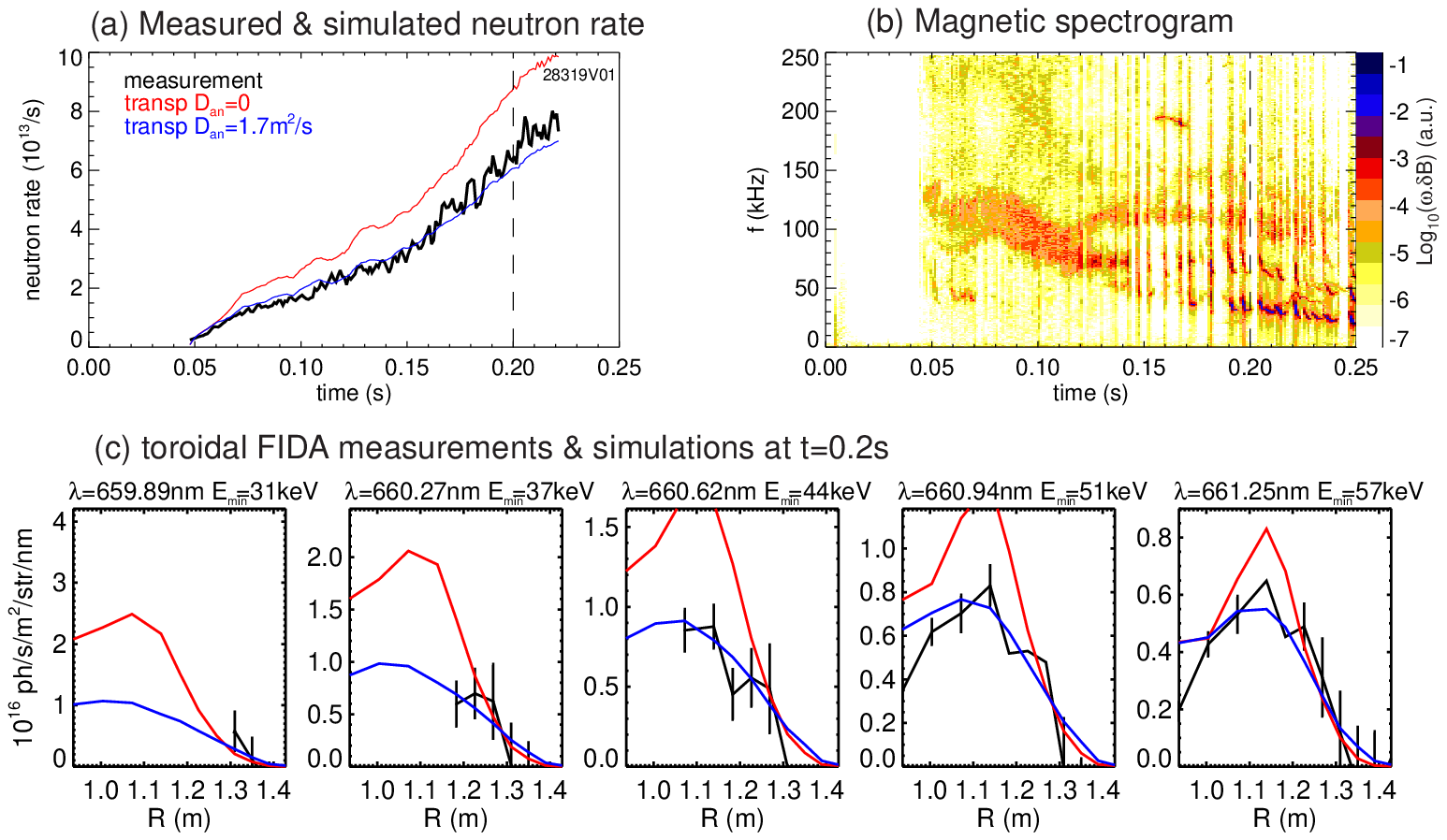}
\caption{Comparison of measured fast ion data with simulations in $\#28319$. (a) Time history of neutron rate and simulation assuming various diffusion models. (b) Magnetic spectrogram showing presence of fishbones and other chirping mode activity. (c) Comparison of measurements with modelling considering different diffusion models late during the discharge (the colours corresponding to particular models are the same as those used in (a)).}
\label{fig:fidatordndfb}
\end{figure*}

FIDA data from a chirping mode discharge $\#28319$ are shown in Figure
(\ref{fig:fidatordndfb}). Initially a 'continuum' of modes exists,
then they become more discretely spaced in time. A uniform anomalous
diffusion coefficient of $1.7\mathrm{m}^2/\mathrm{s}$ brings the
predicted neutron rate into agreement with the fission chamber. This
appears not to require adjustment depending on the period of the
fishbones, reinforcing the idea that the period does not affect the
rate of fast ion redistribution. At $t=0.2$s, the FIDA spatial
profiles of the toroidal channels are compared with the simulations
for zero anomalous diffusion and for $\mathrm{D}_{\rm
an}=1.7\mathrm{m}^2/\mathrm{s}$. Here it is found that the FIDA
measurement is clearly smaller than the classical prediction, but
matches quite reasonably, within error bars, the simulation with a
uniform diffusion coefficient. This is in contrast to the LLM, which
required the diffusion to be concentrated in the core.

\begin{figure}[htb]
\includegraphics[width=8cm]{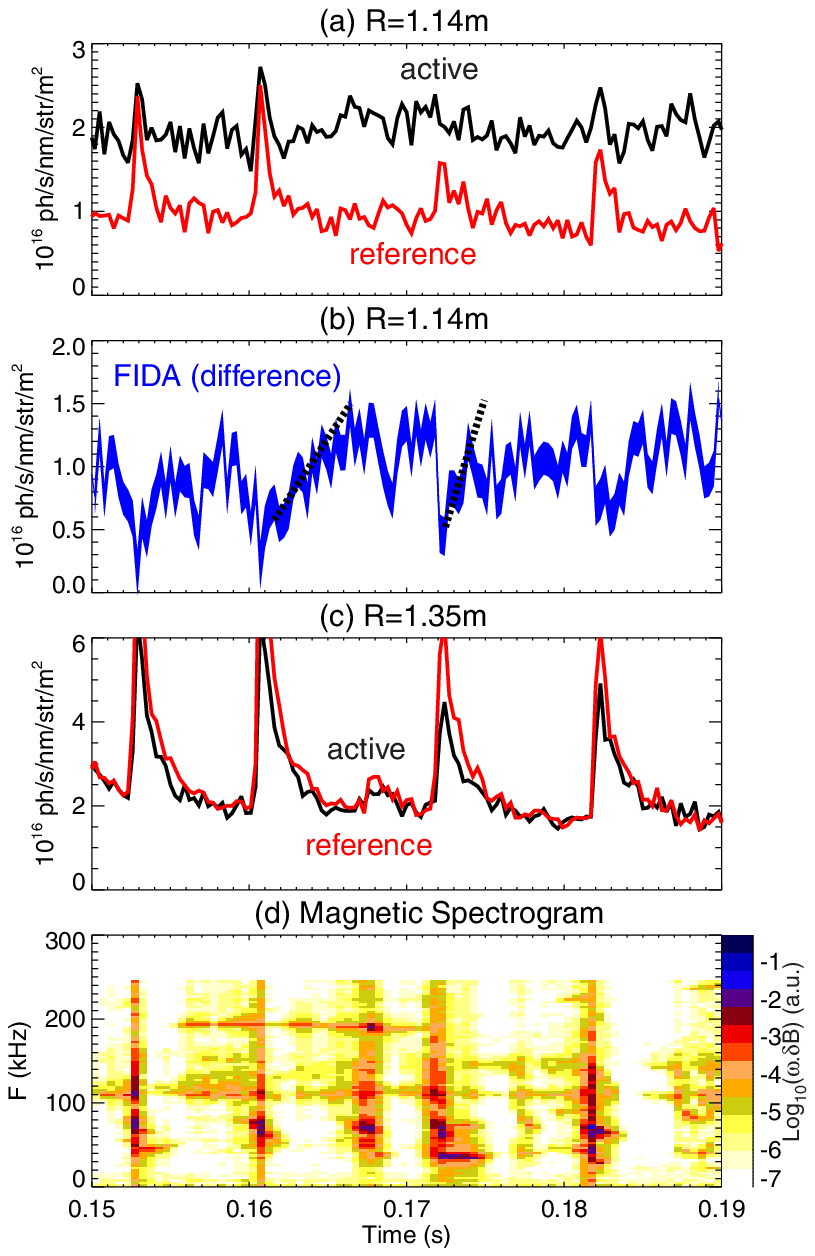}
\caption{Detailed analysis of toroidal FIDA signal evolution during shot $\#28319$ at $\lambda=660.6nm$ ($E_{\min}=44keV$) for a core (a,b) and edge (c) channel, comparing active channel (black) and passive channel (red) in (a,c) as well as the core net active FIDA signal (b), where the thickness of the line denotes the shot noise error bar. (d) Comparison with the magnetic spectrogram.}
\label{fig:chirp}
\end{figure}

To investigate degradation/recovery due to fishbones, the time
evolution of core and edge signals of both active and reference
views (as well as the subtracted, i.e. FIDA component of the core signal), is compared with the magnetic spectrogram in
Figure (\ref{fig:chirp}). This demonstrates the utility of having a
very fast CCD and high light throughput. Clear bursts in the reference signals occur for both core and edge channels, which is considered to be due to a rapid 'burst' of passive FIDA emission from the edge. The core FIDA (difference) signal drops concomitantly with each magnetic burst, with there being some relation between the size of the magnetic burst and the amount of drop of the FIDA signal. There is virtually no difference between active and reference signals at the edge, even during a fishbone burst.  The beam driven FIDA signal is proportional to the beam density ($\sim 10^{15} \mathrm{m}^{-3}$), while the passive FIDA signal is proportional to the neutral density, which can be up to $\sim 10^{18}$ in the edge but attenuates rapidly into the plasma. The large increase in passive FIDA and lack of increase in beam FIDA near the edge indicates that the fast ions are being rapidly redistributed toward the edge. The core (net) FIDA signal on the other hand appears often to recover and 'saturate' at a particular value before the next fishbone occurs.  In fact, one can see a sort of cyclic fast
sawtooth type of behaviour, indicating that smaller bursts may be
occurring in between the larger bursts, and their net transport level
may be similar as the neutron rate does not increase.  Further
simulations of the collisional relaxation between fishbones, as well
as the drop due to individual fishbones, are underway to quantify this
in more detail.

The recovery time of the passive FIDA in the edge is somewhat longer than that of the active signal in the core. It may be that the dynamics of the
recovery of the edge signal are a strong controlling parameter in
triggering the next fishbone.  On the other hand, losses of fast ions at the edge are quite rapid. The mechanism of fast ion damping/orbit loss in the
edge is therefore important.

These fishbone cycles are only prevalent in the toroidal FIDA signals;
the vertical FIDA signals do not show any clear drop with fishbone
events, and although some correlation may be found it is much less
clear. This may be because there is an anomalously low vertical FIDA
signal. Further analysis of FIDA measurements during fishbone activity
in MAST can be found in \cite{owenpaper}.

\section{Conclusions}

The FIDA signals observed on MAST provide both energy and
spatially-resolved information about the fast ion distribution. The
diagnostic arrangement, as well as first results from the dual-view
FIDA spectrometer on MAST have been presented. The toroidal chords can
be used for moderate to high energies (30-60keV) and are strongly
sensitive to passing particles, while the vertical chords can be used
for a broad range of energies and are sensitive to trapped particles
and those close to the trapped/passing boundary (at high
energies). The background signal on toroidal channels is found to
contain passive FIDA emission, driven by edge neutrals, and this can
be almost as large as the active FIDA emission. Moreover, the passive
FIDA emission was found to be quite different in spectral shape and
magnitude depending on which of the views were taken: the active view
through each beam; a toroidally-displaced reference; or a
vertically-displaced reference.  It was concluded that using the
toroidally-displaced reference on the SS beam was most accurate and
good enough for comparing measurements with FIDASIM
simulations. Cross-calibration of all the views has been carried
out. Absolute calibration has been performed, and is partially
validated in that the spectral baseline often agrees with the Zebra
diagnostic. Uncertainties remain, since FIDASIM and NEBULA simulations
predict a factor 1.4 higher intensity beam emission than that measured
with the FIDA diagnostic. Possible causes of this discrepancy have
been considered. Incorporating this factor, the toroidal FIDA signals
have been validated against modelling in the MHD-quiescent phase of a
discharge where the neutron rate also matches simulation. On the other
hand, FIDA signals from vertical chords agree with simulations for
$E_{\rm min} < 11$keV, but do not agree with simulations at higher
energies. This indicates that there may be discrepancies in
simulations (even in the absence of MHD) for pitch angles around the
trapped/passing boundary near the core, which may be due to finite
Larmor radius effects combined with potato orbits in the core. The
NUBEAM simulations simply approximate the gyro-motion as circular
motion in a magnetic field sampled at the guiding centre, which may
not be true in spherical tokamaks where the field varies considerably
around the gyro orbit.  This is a topic for study, and hopefully for
inclusion in future simulations.

FIDA measurements show a degradation in fast ion transport due to the
long-lived mode, and a model which is most consistent with the total
neutron rate and FIDA signal is one in which the anomalous diffusion coefficient is high in the inner half radius and zero outside this. This is consistent with the LLM being localized in the broad, low-shear region of the
core.

In discharges with fishbones, a spatially flat anomalous diffusivity
of a few $m^2/s$, consistent with the neutron rate, is sufficient to
fit the spatial profile of toroidal signals. During fishbones, core
active FIDA signals drop while edge active FIDA signals hardly change
but the passive FIDA signals increase dramatically, indicative of
dramatic sudden redistribution far into the edge, which may be due to
for example an avalanche or induced change of orbit type from confined
to unconfined.  Between fishbones, the active FIDA signal appears to
saturate. There may however be other modes present in the selected
discharge which affect this active FIDA signal. Further work is underway to examine changes in spectra during fishbones and simulate the inter-fishbone evolution. Simulation work is also underway at to quantify fast ion losses by
fishbones using the simulation codes LOCUST-GPU \cite{locust} and HAGIS \cite{hagis}.
Tomographic techniques will also be applied to simultaneous toroidal
and vertical FIDA measurements to generate a reconstruction of the fast ion distribution \cite{mirkotomo}.

\section{Acknowledgments}
The authors would like to acknowledge the support of the MAST
technical team, as well as M. Salewski and K. McClements for checking
the manuscript, A. Bortolon for assistance with the spectrometer,
B. Grierson for information about atomic rate coefficients, and D. Liu
and E. Ruskov for their assistance with the FIDASIM code.
This work was funded partly by the RCUK Energy Programme under grant
EP/I501045 and the European Communities under the contract of
Association between EURATOM and CCFE. The views and opinions expressed
herein do not necessarily reect those of the European Commission or of
the ITER Organization.

\bibliography{references}

\end{document}